\documentclass[aps,amsmath,prb,reprint,superscriptaddress,showpacs,showkeys]{revtex4-1}
\usepackage{graphicx}
\usepackage{color}
\usepackage{ulem}
\newcommand{\red}{\textcolor{red}}

\def\kB{k_{\rm B}}
\def\<{\langle}
\def\>{\rangle}
\def\({\left(}
\def\){\right)}
\def\[{\left[}
\def\]{\right]}

\def\sin{\mathop{\mathrm{sin}}\nolimits}
\def\cos{\mathop{\mathrm{cos}}\nolimits}

\def\kk{\textbf{\textit{k}}}

\def\qq{\textbf{\textit{q}}}
\def\rr{\textbf{\textit{r}}}
\def\jj{\textbf{\textit{j}}}
\def\uu{\textbf{\textit{u}}}

\def\vv{\textbf{\textit{v}}}
\def\xx{\textbf{\textit{x}}}

\def\JJ{\textbf{\textit{J}}}

\def\RR{\textbf{\textit{R}}}

\def\QQ{\textbf{\textit{Q}}}
\def\FF{\textbf{\textit{F}}}
\def\XX{\textbf{\textit{X}}}
\def\RR{\textbf{\textit{R}}}

\begin{document}
\draft
\title{Off-center Rattling Triggers High-Temperature Thermal Transport in Thermoelectric Clathrates: Non-Perturbative Approach
}
\date{\today}

\author{Qing Xi}
\affiliation{Center for Phononics and Thermal Energy Science,\\
School of Physics Science and Engineering, Tongji University, 200092
Shanghai, PR China}
\affiliation{China-EU Joint Lab for Nanophononics, Tongji University, 200092 Shanghai, PR China}
\affiliation{Shanghai Key Laboratory of Special Artificial Microstructure Materials and Technology, School of Physics Science and Engineering, Tongji University, 200092 Shanghai, PR China}

\author{Zhongwei Zhang}
\affiliation{Center for Phononics and Thermal Energy Science,\\
School of Physics Science and Engineering, Tongji University, 200092
Shanghai, PR China}
\affiliation{China-EU Joint Lab for Nanophononics, Tongji University, 200092 Shanghai, PR China}
\affiliation{Shanghai Key Laboratory of Special Artificial Microstructure Materials and Technology, School of Physics Science and Engineering, Tongji University, 200092 Shanghai, PR China}

\author{Tsuneyoshi Nakayama}
\email{tnaka@eng.hokudai.ac.jp}
\affiliation{Center for Phononics and Thermal Energy Science,\\
School of Physics Science and Engineering, Tongji University, 200092
Shanghai, PR China}
\affiliation{China-EU Joint Lab for Nanophononics, Tongji University, 200092 Shanghai, PR China}
\affiliation{Shanghai Key Laboratory of Special Artificial Microstructure Materials and Technology, School of Physics Science and Engineering, Tongji University, 200092 Shanghai, PR China}
\affiliation{Hokkaido University, Sapporo 060-0826, Japan}

\author{Jie Chen}
\affiliation{Center for Phononics and Thermal Energy Science,\\
School of Physics Science and Engineering, Tongji University, 200092
Shanghai, PR China}
\affiliation{China-EU Joint Lab for Nanophononics, Tongji University, 200092 Shanghai, PR China}
\affiliation{Shanghai Key Laboratory of Special Artificial Microstructure Materials and Technology, School of Physics Science and Engineering, Tongji University, 200092 Shanghai, PR China}

\author{Jun Zhou}
\email{zhoujunzhou@tongji.edu.cn}
\affiliation{Center for Phononics and Thermal Energy Science,\\
School of Physics Science and Engineering, Tongji University, 200092
Shanghai, PR China}
\affiliation{China-EU Joint Lab for Nanophononics, Tongji University, 200092 Shanghai, PR China}
\affiliation{Shanghai Key Laboratory of Special Artificial Microstructure Materials and Technology, School of Physics Science and Engineering, Tongji University, 200092 Shanghai, PR China}

\author{Baowen Li}
\email{Baowen.Li@colorado.edu}
\affiliation{Department of Mechanical Engineering, University of Colorado, Boulder, Colorado 80309, USA}


\begin{abstract}
 
Lattice thermal conductivities $\kappa_{\rm L}$ of type-I clathrates encapsulating ``off-center" guest atoms
behave in a way that is sharply different from those of conventional crystals such as on-center type-I clathrates.
The latter decreases with increasing temperature according to 1/$T$ above a few tens\,K.
While, $\kappa_{\rm L}$ of off-center clathrates increase T-linearly  above a few tens\,K.
Further increasing temperature above $T\gtrsim$100\,K, $\kappa_{\rm L}$ saturates without exhibiting appreciable T-dependence. 
In this temperature regime, 
the standard formulation of $\kappa_{\rm L}$ based on the perturbation theory is unfeasible.
We herein present a theory of  $\kappa_{\rm L}$
based on ``non-perturbative" approach by taking into account the interaction between non-vibrational off-center rattling states of guest atoms and cage shells.
It is remarkable that our theory shows excellent agreement with observed $\kappa_{\rm L}(T)$ for off-center clathrates in magnitudes, the cage volume $\Omega$-dependence and  T-dependence.

\end{abstract}

\maketitle

\section{INTRODUCTION}
High-temperature thermal transport is a crucial issue in the search of efficient thermoelectric materials.
This is because operating temperatures of thermopower devices are higher than room temperature.
Efficient thermoelectric effects should be achieved by materials possessing both ``glass-like" low lattice thermal conductivity and crystal-like high electrical conductivity.\cite{Slack1995,Beekman2015,Takabatake2014}
Here, the glass-like thermal conductivity is characterized from low temperature regions as follows\,\cite{Takabatake2014}: (i)  $\kappa_{\rm L}$ is proportional to $T^{2+\delta}$ ($\delta\ll 1$)below a few K.
(ii) The plateau region from a few K to a few tens\,K.
(iii) The T-linear rise above the plateau from $\sim$10 K to 100\,K.
(iv) The saturate regime independent of $T$ above $T\gtrsim$100\,K.
These are the prominent  hallmarks of ``glass-like" thermal conductivities, almost identical to those of structural glasses.

Type-I clathrates are promising thermoelectric material fulfilling the concept of phonon-glass electron crystal concept. In actual, $\kappa_{\rm L}$ of type-I clathrates with off-center guest atoms behave like those of structural glasses.\cite{Takabatake2014,Beekman2015} See Fig. 1.
Type-I clathrate R$_{8}$M$_{16}$Z$_{30}$ (R=Ba, Sr, Eu; M and Z=Al, Ga, In, Si, Ge, Sn) has a $Pm\bar{3}n$ cubic structure,  formed by the network cages consisting of face-sharing polyhedrons (dodecahedron and tetrakaidecahedron) and the encapsulated guest atoms.\cite{Takabatake2014} 
Guest atoms in tetrakaidecahedral cages will take off-center positions when the ionic radius of guest atoms is relatively small compared with the cage radius\,\cite{Sales2001}, which has been observed in Sr$_{8}$Ga$_{16}$Ge$_{30}$\,\cite{Chakoumakos2000, Sales2001}, Eu$_{8}$Ga$_{16}$Ge$_{30}$\,\cite{Sales2001}, Ba$_{8}$Ga$_{16}$Sn$_{30}$\,\cite{Suekuni2008}, Ba$_{7.81}$Ge$_{40.67}$Au$_{5.33}$\,\cite{Lory2017} by means of x-ray and neutron diffraction methods.
The off-center guest atoms break the translational invariance of the systems. 
More importantly,  the weakly bounded guest atoms in cages with large open space execute non-vibrational motion above $T\gtrsim$100K with large atomic displacement parameter.\cite{Chakoumakos2001, Qiu2004, Okamoto2006, Christensen2013, Chakoumakos2000, Chakoumakos2001}
In this context, we should mention the concept of part-crystalline part-liquid state which has been proposed for filled skutterudites,\cite{Qiu2014,Qiu2016} indicating that the non-vibrational state of rattling guest atoms is a general aspect at high temperature for host-guest systems.
\begin{figure}[t]
\includegraphics[width=1.0\linewidth]{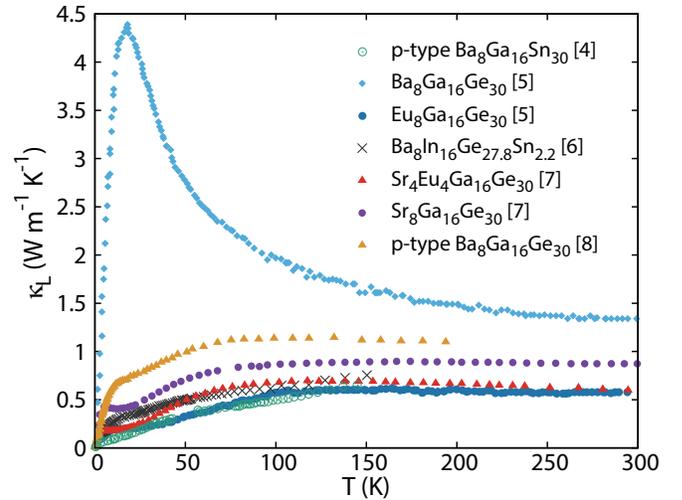}
\caption{(Color online)
Temperature dependence of $\kappa_{\rm L}$ plotted in linear scale for
various off-center type-I clathrates\,\cite{Suekuni2008, Suekuni2007,Cohn1999,Bentien2004,Sales2001} and ``on-center" type-I $\rm Ba_{8}Ga_{16}Ge_{30}$\,\cite{Sales2001}. 
The parentheses [...] in inset give reference numbers.
}
\label{fig61}
\end{figure}

Let us begin by giving a brief summary on previous theoretical results on $\kappa_{\rm L}(T)$ of off-center type-I clathrates.
The (i) described in the first paragraph in Introduction has been successfully interpreted by the two-level tunnelling states caused from  frustrated electric-dipole configuration concerning to off-center guest ions.\cite{Nakayama2008, Kaneshita2009}
The relevance of the localization-delocalization (LD) transition  on the onset of the plateau at a few K\,(ii)  has been demonstrated for the first time by means of large-scale numerical simulations.\cite{Liu2016}
In addition, the T-linear rise above a few tens\,K (iii) has been explained by the hopping of strongly localized modes assisted by acoustic phonons.\,\cite{Xi2017}
For these disordered systems, large-scale simulation based on \textit{ab initio} calculation is difficult as seen from the discrepancy between simulation results and experimental measurements of room temperature lifetimes.\cite{Lory2017} 
While, the coarse-grained Hamiltonian is effective since \red{the} wave-lengths $\lambda$ of acoustic phonons at THz frequency region are much larger than the size of unit cell $a_0\simeq$1\,nm including 8 guest atoms and 46 cage atoms \red{bounded} by strong covalent-bonding.
However, despite the importance of the issue (iv) in the search of efficient thermoelectric materials, the underlying mechanism is yet to be explained.
Here we aim at a theoretical elucidation of the underlying mechanism of heat transfer at $T\gtrsim$100\,K. 
For this purpose, we discuss the problem from a general point of view by proposing a theoretical model of heat transfer in host-cage systems with off-center rattling guest atoms. In this sense, the mechanism proposed here is applicable to a wide class of host-guest systems involving weakly bounded atoms in void/open structures.

\section{Energy flux in the rattling regime  $T\gtrsim T\rm{_R}$}
Type-I clathrates encapsulating ``off-center" guest atoms in tetrakaidecahedron cages are composed of networked host-cages of a simple cubic structure. 
We give a 3D Video obtained from molecular dynamics (MD) simulations for  type-I $\rm Ba_{8}Ga_{16}Sn_{30}$ involving off-center guest atoms in Supplemental Material\,\cite{SM_Video}, which visualizes that guest atoms at 400\,K perform ``off-center" rattling motion, which are not assigned as vibrational states satisfying the condition $\omega\tau>1$.
Meanwhile, cage atoms vibrate with smaller amplitudes around their equilibrium positions described as vibrational states.
At 10\,K, however, guest atoms locate at off-center positions in cages, and vibrate there. 

This aspect is consistent with THz-frequency spectroscopy data for off-center type-I BGS
observed at various temperatures.\,\cite{Mori2011, Kume2015}
The data exhibit two peaks at 0.5 and 0.7 THz frequencies at $T_{\rm R}\simeq$100\,K.
Those are assigned as librational and stretching modes relevant to off-center guest atoms in type-I BGS.\,\cite{Nakayama2011}
Increasing temperature, these two peaks collapse into a single peak at 0.75 THz with comparable width. 
However, such a characteristic has not been observed for on-center type-I BGG\,\cite{Mori2011, Kume2015}.
These spectroscopic data provide a strong evidence for off-center guest atoms to behave as off-center rattlers  in cages at $T\gtrsim T\rm{_R}$.

The rattling temperature $T\rm{_R}$ should play a crucial role for $\kappa_{\rm L}(T)$ in type-I clathrates with off-center guest atoms. 
Firstly, at $T\lesssim T\rm{_R}$, the systems take disordered configuration of guest atoms, and  it converts into higher symmetry at $T\gtrsim T\rm{_R}$ where guest atoms behave as rattlers in cages. 
Secondary, the strong covalent-bonding between cage atoms ensures the existence of acoustic phonons in host cages even at $T\gtrsim T\rm{_R}$ in contrast to rattling states of weakly bound guest atoms\,\cite{SM_Video}.
At $T\lesssim T\rm{_R}$, the standard kinetic formula $\kappa_{\rm L}(T)=1/(3V)\sum_{\bf k{\rm s}}C_V(\omega_s)v_s^2\tau(\omega_s)$
is valid under the condition $\omega\tau>1$\cite{Landau_Physical_Kinetics}.  
At $T\gtrsim T_{\rm R}$, however, off-center guest atoms cannot be described as vibrational states and 
the perturbation treatment is unavailable in contrast to the case of on-center clathrates.\cite{Tadano2015, Tadano2018} 

We employ the formulation of energy flux available for non-vibrational states of rattling atoms.
Hardy\,\cite{Hardy1963} has  presented the rigorous formulation on the heat transfer, from which it is straightforward to deduce the version of energy flux working at  $T\gtrsim T\rm{_R}$  by taking the limit $\hbar\rightarrow$ 0.
The local heat flux density $\jj(\xx, t)$ is obtained from the continuity equation for the  energy density $h(\xx,t)=\sum_{\ell}\epsilon_\ell(t)\delta(\xx-\rr_\ell)$ given by the time-dependent total energy $\epsilon_\ell (t)$  of $\ell$-th atom at the position vector $\rr_\ell$. 
The average energy flux is defined as $\JJ=(1/V)\int^V\jj(\xx)dV$, which yields\,\cite{Voltz2000}
\begin{eqnarray}
\JJ(t)=\frac{1}{V}\frac{d}{dt}\left[\sum_\ell\rr_\ell (t) \epsilon_\ell (t)\right] .
\label{eq_JJ_11}
\end{eqnarray}
By decomposing the contributions from guest atom $i$ and cage atom $j$ by taking $\ell=i~{\rm or}~j$, we have the following expression for the guest atom energy $\varepsilon_i(t)$ and the cage-atom's $E_j(t)$, which are composed of  the interaction potentials $V_{ij}^{CG}$  between guest atoms ($i)$ and cage atoms ($j$) and $V_{jj'}^{CC}$ between cage atoms ($j,j'$), 
\begin{eqnarray}
\varepsilon_i(t)&=&\frac{1}{2}m\dot{\xx}_i^2(t)+\frac{1}{2}\sum_jV_{ij}^{CG}(\XX_j,\xx_i),
\nonumber
\\
E_j(t)&=&\frac{1}{2}M_j\dot{\XX}_j^2(t)
+\frac{1}{2}\sum_{j'}V_{jj'}^{CC}(\XX_j,\XX_{j'})
\nonumber
\\
&+&\frac{1}{2}\sum_iV_{ji}^{CG}(\XX_j,\xx_i),
\label{eq_EE_11}
\end{eqnarray}
where $m$ and $M_j$ are masses of guest atoms and cage atoms,
and the subscript $j$ on cage atoms is combined with the subscript $i$ of a guest atom since each cage involves only one guest atom, $i.e.$, $j\in i$. 
The position vector $\xx_i(t)$ of $i$-th guest atom is expressed by the sum of the small displacement $\qq_i(t)$ and the equilibrium position $\RR_i$, $\xx_i(t)=\RR_i+\qq_i(t)$, and the position vector of cage atom $j$ is given by  $\XX_j(t)=\RR_j+\QQ_j(t)$ with  $\QQ_j(t)$ the small displacement of cage atom.
$\dot{\xx}_i(t)$ and $\dot{\XX}_j(t)$ are time-dependent velocities of the $i$-th guest atom and the $j$-th cage atom, respectively. 

The substitution of Eq.\,(\ref{eq_EE_11}) into Eq.\,(\ref{eq_JJ_11}) leads to\,\cite{Hardy1963, Voltz2000}
\begin{eqnarray}
\JJ(t)
&=&\frac{1}{V}\sum_{i,j\in i}\left[\varepsilon_i(t)\dot{\xx}_i(t)+E_j(t)\dot{\XX}_j(t) \right]
\nonumber
\\
&+&\frac{1}{V}\sum_{j,j'}\left[\left(\XX_j(t)-\XX_{j'}(t)\right)\FF_{jj'}^{CC}\cdot\dot{\XX}_j(t)\right]
\nonumber
\\
&+&\frac{1}{V}\sum_{i,j\in i}\left[\left(\xx_i(t)-\XX_j(t)\right)\FF_{ij}^{GC}\cdot\left(\dot{\xx}_i(t)+\dot{\XX}_j(t)\right)\right],
\label{eq_JJ_22}
\end{eqnarray}
where the interatomic force  $\FF_{jj'}^{CC}$ between cage atoms $j, j'$ and $\FF_{ij}^{GC}$ between guest atom $i$ and cage atom $j$ are defined by
\begin{eqnarray}
\FF_{jj'}^{CC}=-\frac{\partial U_{j'}^{CC}}{\partial \XX_j}, \FF_{ij}^{GC}=-\frac{\partial U_{j}^{GC}}{\partial\xx_i},
\end{eqnarray}
with
\begin{eqnarray}
U_{j}^{CC}=1/2\sum_{j'}{V_{jj'}^{CC}(\XX_j,\XX_{j'})},
\nonumber
\end{eqnarray}
and
\begin{eqnarray}
U_{j}^{GC}&=&1/2\sum_{i}{V_{ji}^{GC}(\XX_j,\xx_i)},
\end{eqnarray}
where $U_{j}^{CC}$ is the cage atom-cage atom  interaction potential for the cage atom $j$, and $U_{j}^{GC}$ is the guest-cage potential for the cage atom $j$.
Equation (\ref{eq_JJ_22}) is applicable for non-vibrational states in gases and liquids.
The first parenthesis in the right in Eq. (\ref{eq_JJ_22}) describes the convective contribution to heat transfer, and the second and the third in the right are the conductive part. 
In off-center type-I clathrates, the location of guest atoms deviates significantly from the off-center equilibrium position at $T\gtrsim T_{\rm R}$. 
It is apparent, however, that guest atoms cannot diffuse to nearest neighbour cages. 
Consequently, we do not take into account the convectional terms in Eq. (\ref{eq_JJ_22}). 

The second parenthesis in the right in Eq. (\ref{eq_JJ_22}) expresses the energy transfer in host cage, whose constituent atoms are bound by short-range covalent bonding.
Type-I clathrates consist of 8 guest atoms and 46 cage atoms in unit cell, where optical modes attributing to these atoms are irrelevant to heat transfer.
In addition, the time average of the relative distance between cage atoms in the same cage becomes 
$\overline{\XX_j(t)-\XX_{j'}(t)}=d_{jj'}$ for the dominant nearest neighbour coupling, where $d_{jj'}$ is the nearest-neighbour distance due to short-range covalent bonding, which is much smaller than the size of cage $2R$.
While, at $T\gtrsim T_{\rm R}$, the relations 
$\overline{\xx_i-\XX_j}\cong R$ and $\dot{\xx}_i(t)\gg\dot{\XX}_{j\in i}$ holds in the third parentheses\,\cite{SM_Video}.
As a result, 
the third term stemmed from rattling guest atoms should play a key role for the  heat transfer at $T\gtrsim T{\rm _R}$. 
Thus, the dominant term for energy flux at $T\gtrsim T{\rm _R}$ in Eq. (\ref{eq_JJ_22}) is given by
\begin{eqnarray}
\JJ(t)=\frac{1}{V}\sum_{i,j\in i}^N\left[\xx_i(t)-\XX_j(t)\right] \FF_{ij}\cdot\dot{\xx}_i(t).
\label{eq_FF_33}
\end{eqnarray}
To assess more definitely the impact of guest atoms on energy flux, we consider another aspect of the part of $\FF_{ij}\cdot\dot{\xx}_i$  in Eq. (\ref{eq_FF_33}) using the relation 
\begin{equation}
\sum_{j\in i}\FF_{ij}\cdot\dot{\xx}_i(t)=\dot{\varepsilon}_{i}.
\end{equation}
This relation yields the  thermal/time-averaged energy flux (heat flux) at $T$ given by
\begin{eqnarray}
\langle\overline{\JJ(t)}\rangle=
\frac{N}{V} R\dot{\varepsilon}_{c}(T),
\label{eq_FF_44}
\end{eqnarray}
where $\langle\cdots\rangle$ denotes the thermal average at a temperature $T$. 
The energy fluxes from individual cages are identical at each site because of the translational invariance of the system at the rattling regime $T\gtrsim T_{\rm R}$.
So, we have employed the notation $\varepsilon_c$ for $\varepsilon_{i}$.  

Let us give a qualitative interpretation on the above formula. 
The kinetic energy $\varepsilon_c$ can be related with the local
pressure $p_c$ in a cage and the cage volume $v_c$ by the relation  $\varepsilon_c=p_c v_c$. 
The volume change $v_c+\delta v_c$  caused by the interaction between guest atom and cage atoms yields the energy change  $\delta \varepsilon_c(t)=\varepsilon_c\Delta_c(t)$, where  the dilation is given by  $\Delta_c (t)=\delta v_c/v_c$. 
Consequently, the off-center rattling motion of guest atom trigger the energy transfer to host cage via deformation of cage shells. 
In thermal equilibrium, guest atoms convert their energies into cage-shell modes, and those are recovered by the reverse process.
 

\section{Energy flux from rattling guest atoms to host cage at $T\gtrsim T_{\rm R}$}
Off-center rattling motion of guest atoms at $T\gtrsim T_{\rm R}$ triggers the energy exchange with host cages  via the deformation of cages. 
This mechanism is identical to the emission of phonons from  a small particle immersed in an elastic medium.
We utilize the treatment on heat transfer from fine particles immersed in liquid He II\,\cite{Nakayama1981} and liquid $^3$He\,\cite{Nishiguchi1982} by modifying those to be applicable at high temperatures $T\gtrsim T_{\rm R}$.
The mean rate of energy emitted from dilation modes of a cage is given by the square of the velocity field $\vv(\rr,t)$\,\cite{Landau_Fluid},
\begin{eqnarray}
\dot{\varepsilon}_{\rm c}=\rho_{\rm s} v_{\rm \parallel}\oint_S |\vv(\rr)|^2dS,
\label{eq77}
\end{eqnarray}
where $\rho_{\rm s}$ and $v_{\rm \parallel}$ are the mass density and the  velocity of longitudinal phonons in host cage.
$dS$ is the surface element surrounding a cage.
Since the emitted energy is carried away by vibrations of a cage shell, it needs to relate the velocity field $\vv(\rr)$ at $\rr$ to $\vv(\RR)$ at the surface $\RR$ under the situation $|\rr|\geq|\RR|$.

The velocity fields at $\rr$ and $\RR$ can be related by introducing the scalar potential $\phi_q(\rr,t)$ defined by $\vv(\rr,t)={\rm grad}\,\phi(\rr,t)$\,\cite{Nakayama1981}.
The radial component of the velocity field becomes, 
\begin{eqnarray}
|\vv^q(\rr)|^2=\frac{R^4}{r^4}\frac{1+(qr)^2}{1+(qR)^2}|v_r^q(\RR)|^2,
\label{eq99}
\end{eqnarray}
where $q$ represents the wave number of emitted phonon.
This provides the emission of spherical pressure (longitudinal) waves\,\cite{Nakayama1981}.
The effect of shear (transverse) waves stemmed from the angular components $v_\theta^q(R)$ and $v^q_\phi(R)$ will be discussed later.

Substituting this relation into Eq.\,(\ref{eq77}) and taking the closed surface of the integration to be a sphere of radius $r> R$, the mean energy emitted from a cage becomes
\begin{eqnarray}
\dot{\varepsilon}_c^q=\rho_{\rm s}v_{\parallel}\frac{q^2R^4}{1+q^2R^2}\int_0^{2\pi}\int_0^{\pi}|v_r^q(\RR)|^2d\Omega,
\label{eq99}
\end{eqnarray}
where $d\Omega=\sin\theta d\theta d\phi$.
The velocity field $v_r(\RR)$ at the surface of a cage must equal to the time derivative of the surface displacement of the cage, $i.e.$, $\vv(\RR,t)=\dot{\uu}(\RR,t)$.
The $\uu(\RR,t)$ is expressed by the sum of the eigenmode $J$ for shell vibrations of cage, where  $J$ stands for a set of quantum numbers ($\ell,m,\omega_q^{\ell,m}$) specifying the eigenmodes.
By defining the eigenmode belonging to the eigenfrequency $\omega_J$ as $\vec{\psi}_J(\RR)$,
the displacement operator is expressed by 
\begin{equation}
\uu(\RR,t)=\sum_J\sqrt{\frac{\hbar}{2\rho_{\rm s}\omega_J\Omega}} (a_J\vec{\psi}_J(\RR)e^{-i\omega_J t}+\rm{H.c.}),
\end{equation}
where $\Omega$ and $\rho_{\rm s}$ are the volume and the mass density of cage, respectively, and  $a_J$ and $a_J^\dagger$ are the Boson operators of the mode $J$.
The eigenfunction $\vec{\psi}_J(r)$ represents spherical wave emitted from a cage.

The thermal-averaged total energy emitted from a cage via longitudinal phonons is obtained by summing up $J$ for Eq.\,(\ref{eq99}), which yields
\begin{eqnarray}
&\dot{\varepsilon}_c(T)=\rho_{\rm s}v_{\rm \parallel}\sum_J\frac{q^2R^4}{1+q^2R^2}
\int_0^{2\pi}\int_0^{\pi}d\Omega
\nonumber
\\
&\times<\dot{u}_r^J(\RR,t)^\dagger\dot{u}_r^J(\RR,t)>,
\label{eq1111}
\end{eqnarray}
where $\dot{u}_r(\RR)$ represents the normal component of the velocity field
 at the surface of a cage.
The thermal average is defined by 
\begin{equation}
<\dot{u}_r^J(R)^\dagger \dot{u}_r^J(R)>={\rm Tr}[e^{-\beta H}\dot{u}_r^J(R)^\dagger \dot{u}_r^J(R)],
\end{equation}
where $H=\sum_J\hbar\omega_J a_J^\dagger a_J$ and $\beta=1/k_{\rm B}T$.
Noting $<a_J^{\dagger} a_J>=n(\omega_J,T)$ given by the Bose-Einstein distribution function, the integral in Eq.\,(\ref{eq1111}) yields\,\cite{Nakayama1981}
\begin{eqnarray}
&\int_0^{2\pi}\int_0^{\pi}<\dot{u}_r^J(\RR)^\dagger\dot{u}_r^J(\RR)>d\Omega
=\frac{\hbar\omega_J}{\rho_{\rm s}\Omega}n(\omega_J,T)
\nonumber
\\
&\times\int_0^{2\pi}\int_0^{\pi}
d\Omega
\left(\frac{\partial j^\ell}{\partial x}\mid_{r=R}^2+\left[ {\rm Higher\,order\,terms}\right]\right),
\label{eq1212}
\end{eqnarray}
where $j^\ell(x)$ with $x=qr$ is the $\ell$-th order spherical Bessel function of the first kind.
\begin{figure}[t]
\includegraphics[width=1.0\linewidth]{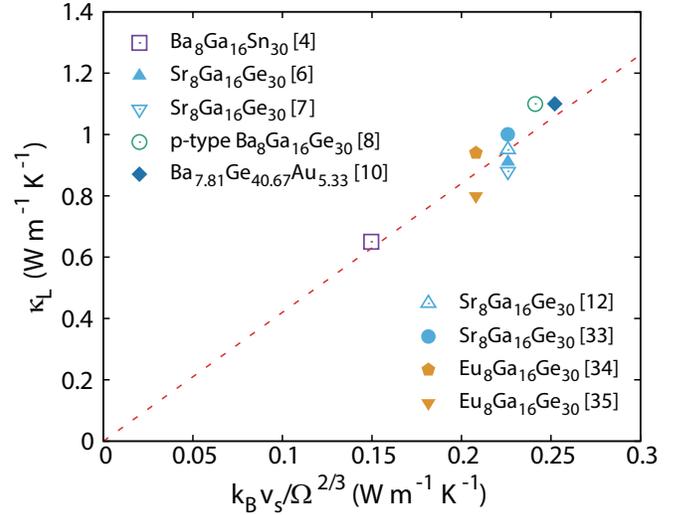}
\caption{(Color online) Observed $\kappa_{\rm L}$ for various off-center type-I clathrates as a function of $k_{\rm B}v_{\rm s}/\Omega^{2/3}$\,\cite{Suekuni2008, Suekuni2007, Cohn1999, Bentien2004, Lory2017, Qiu2004, Christensen2016, Paschen2001, Bentien2005}.
Solid inverted-triangle on off-center Eu$_{8}$Ga$_{16-x}$Sn$_{30+x}$ denotes the thermal conductivity averaged for $x=0.47-0.76$\,\cite{Bentien2005}.
The dashed line is a guide for the eyes showing good agreement with Eq.\,(\ref{eq1416}) for $\gamma=4.2$.
The parentheses [...] in inset give reference numbers.
}
\label{fig66}
\end{figure}
It should be emphasized that, since the size of cage is small enough 
$2R\approx 5\AA$, the lowest symmetric mode of cavity-cage dominantly contributes to the energy transfer.
Higher order terms in the parenthesis in Eq.\,(\ref{eq1212}) provide small contribution.
Thus, the most effective process occurs in the fundamental mode ($\ell$=0) with the eigenfrequency $\omega^0=2\pi v_s/\lambda_0$ attributing to the lowest spherical Bessel function $\ell$=0 of the form $j^0(x)=\sin x/x$ in Eq.\,(\ref{eq1212}).
The wave-length $\lambda_0$ should be $\lambda_0=4R$ for the cage with the radius $R$.

The heat conductance is defined by $h(T)=\partial_TJ(T)$ from the relation $J(T+\Delta T)-J(T)=h(T)\Delta T$.
By taking the high temperature relation $n(\omega_J,T)=k_{\rm B}T/\hbar\omega_J$ in Eq.\,(\ref{eq1212}) and 
substituting the results of Eq.\,(\ref{eq1111}) into Eq.\,(\ref{eq_FF_44}), we have the heat conductance due to the emission of longitudinal acoustic phonons from a cage as 
\begin{eqnarray}
h_{\parallel}(T)=
\frac{4\pi k_{\rm B}v_{\rm \parallel}q_o^2R^5}{(1+q_0^2R^2)\Omega^2}
\frac{\partial j^0(q_0r)}{\partial x}\mid_{r=R}^2,
\label{eq1313}
\end{eqnarray}
where we have used the relation $N/V=1/\Omega$ for the number density of off-center guest atoms, not the number density of all atoms in the systems.
It should be noted that $\hbar$ is cancelled out in Eq.\,(\ref{eq1313}).
This is consistent with the fact that the classical-dynamics works in the rattling regime $T\gtrsim T_{\rm R}$.  
The $\partial j^0(x)/\partial x\mid_{r=R}$ in Eq.\,(\ref{eq1313}) equals the 
first-order spherical Bessel function
\begin{equation}
j^1(x)=\sin x/x^2-\cos x/x,
\end{equation} 
with $x=q_0R=\pi/2$ from $4R=\lambda_0$.
This gives rise to the heat conductance
\begin{equation}
h_{\parallel}(T)=\alpha_{\parallel} k_{\rm B}v_{\parallel}/\Omega,
\end{equation}
where the prefactor is given by
\begin{equation}
\alpha_{\parallel}= 3(q_0R)^2j^1(q_0R)^2/(1+q_0^2R^2),
\end{equation} 
using the relation $\Omega=4\pi R^3/3$.
It is straightforward to incorporate the effect of shear (transverse $\perp$) waves 
stemmed from the angular components of the velocities $v_\theta^q(\RR)$ and $v^q_\phi(\RR)$ according to the treatment of the heat transfer from small particles to liquid $^3$He\,\cite{Nishiguchi1982}.
Note that transverse modes propagate in liquid $^3$He, but not in liquid $^4$He.
The prefactor $\alpha_\perp$ is obtained by taking into account the dominant contribution from the lowest order of modes, $i.e.$, by replacing $\partial j^0(x)/\partial x\mid_{r=R}^2$ and $v_\parallel$ in Eq.\,(\ref{eq1313}) with  the eigenfunctions of shear modes and $v_\perp$\,\cite{Nishiguchi1982}.

The thermal conductivity $\kappa_{\rm L}$ is obtained from the total conductance $h(T)=h_\parallel(T)+h_\perp(T)$ by multiplying 
the lattice constant $a_0$ of type-I clathrates with off-center guest atoms using the relation $a_0/2=2R$.
This is expressed in terms of the cage volume $\Omega$ as, 
\begin{eqnarray}
\kappa_{\rm L}(T)=\gamma\frac{k_{\rm B}v_{\rm s}}{\Omega^{2/3}},
\label{eq1416}
\end{eqnarray}
where $v_{\rm s}$ is the average velocity of acoustic phonons defined by $v_{\rm s}=(v_{\parallel}+2v_{\perp})/3$. 
This expression directly provides the relevance of guest atoms to high-temperature $\kappa_{\rm L}(T)$  via the cage volume $\Omega$.
We can evaluate the prefactor as $\gamma=2.61$ from the expressions of $\alpha_\parallel$ and $\alpha_\perp$.
This $\gamma$-value attributes to the contribution from fundamental modes of a cage, so it should be considered as the lower limit of $\kappa_{\rm L}^{min}$ at $T\gtrsim T_R$.
The proper value of $\gamma$ can be obtained by incorporating higher-order terms of modes.
Thus, the prefactor $\gamma=4.2$ estimated in Fig.\,\ref{fig66} can be recovered.
Equation\,(\ref{eq1416}) shows T-independent saturated $\kappa_{\rm L}$ in accord with the observations shown in Fig.\,\ref{fig61}.
The $\Omega$-dependence in Eq.\,(\ref{eq1416}) is consistent with the experimental data for off-center clathrates showing ``glass-like" thermal conductivity as seen from Fig.\,\ref{fig66}, in addition to the agreement with observed magnitudes of $\kappa_{\rm L}(T)$.

\section{Relevance to the concept of the minimum thermal conductivity}
The concept of the minimum thermal conductivity (MTC)
was first proposed by Slack\,\cite{Slack1979} by employing the standard formula of thermal conductivity $\kappa(T)=(1/3V)\sum_{\kk}C_Vv_s\ell$.
This formula yields 
\begin{equation}
\kappa_{min}^{SK}(T)=\frac{\kB v_s\left(6\pi^2n\right)^{2/3}}{2\pi}\times I(x_D),
\label{eqslack}
\end{equation}
where the transport integral is given by 
\begin{equation}
I(x_D)=\frac{2}{x_D^2}\int_0^{x_D}\frac{x^3 e^x dx}{(e^x-1)^2}.
\end{equation}
Here, $x=\hbar\omega/\kB T$ and $x_D=\hbar\omega_D/\kB T$ with the Debye cut-off frequency $\omega_D$.
The number density of atoms $n$ is related to the Debye temperature by $\theta_D=\hbar v_s (6\pi^2 n)^{1/3}/\kB$ of "monoatomic" systems. 
The term $x^3$ in the transport integral comes from the combination of the Debye density of states and the postulated mean-free-path of acoustic phonons of the form $\ell(\omega)=2\pi v_s/\omega$ or equivalently 
$q\ell(\omega)=2\pi$, which is identical to the Ioffe-Regel (IR) criterion for strongly localized (SL) modes except the factor 2. 
In the SL regime, the mean-free-path $\ell$ and the localization length collapse into a single length scale, $i.e.$, those have the same meaning.
The prefactor  $\kB v_s\left(6\pi^2n\right)^{2/3}/2\pi$ in $\kappa_{min}^{SK}(T)$ provides the MTC  at high temperatures.

Cahill and Pohl\,\cite{Cahill1987}
have explicitly postulated
that ``all" acoustic phonons are strongly localized  with the localization length $\ell\simeq\lambda/2$ according to the Ioffe-Regel (IR) criterion
of $q\ell(\omega)=\pi$\,\cite{Ioffe1960}. 
The formula\,\cite{Cahill1987} is given by 
\begin{equation}
\kappa_{min}^{CP}=\frac{\kB v_s(6\pi^2 n)^{2/3}}{4\pi}\times I(x_D).
\end{equation}
This is the same as  $\kappa_{min}^{SK}$ in Eq.\,(\ref{eqslack}) except the factor 2 in the denominator stemmed from the IR criterion employed.
These $\kappa_{min}^{SK/CP}$ are proportional to the number density of atoms according to $n^{2/3}$.
In applying the formula for actual systems with many atoms in unit cell such as the case of type-I clathrates with 54 atoms in unit cell or fullerene derivatives\,\cite{Duda2013, Wang2013}, 
it is crucial that the Debye model\,\cite{Cahill1987, Slack1979} not allowing modes $\omega\geq\omega_D$ exhausts the number of degrees of freedom. 
Thus, the MTC formula gives rise to the overestimation of $\kappa_{min}$ for these systems.\,\cite{Duda2013, Wang2013}
In addition, the formulae\,\cite{Cahill1987, Slack1979} give the temperature dependence proportional to $T^2$ at $T\lesssim\theta_D$.
This prediction is conflict with the experimental data $\kappa_L (T)\propto T$ above the plateau temperature region from a few tens\,K to $\sim$100K of off-center clathrates and structural glasses (iii).\cite{Takabatake2014}
The discrepancy arises from their postulate that all acoustic phonons are strongly localized.
This postulate is ``only" valid for fracton excitations in fractal structures.\cite{Nakayama1994}
Thus, the physical basis of our theory based on non-perturbative approach is distinct from these MTC. 

\section{Summary}
We have given a theory on the underlying mechanism governing high-temperature $\kappa_{\rm L}(T)$ of type-I clathrates with off-center guest atoms, 
focusing on the regime at $T\gtrsim T_R$.
Our approach is based on non-perturbative treatment, different from the standard one according to the perturbation theory. 
As seen from the  3D Video of MD simulations in Supplemental Material, off-center rattling motion of guest atoms perform spontaneous displacement with large amplitude in cages, which induces the deformation of surrounding cages, and the energy exchange between cage and guest atom occurs via phonon emission/absorption. 
We point out that the theory proposed here is applicable for host-guest systems involving weakly bound atoms in void/open structures, although we have exemplified the systems of off-center clathrates.
It is remarkable that the prediction of the formula Eq.\,(\ref{eq1416}) shows excellent agreement with observed $\kappa_{\rm L}(T)$ for off-center clathrates
in magnitude, temperature dependence and the cage-volume $\Omega$ dependence.

To conclude, our theoretical investigation based on non-perturbative approach
provides one of  guiding principles for searching and designing efficient thermoelectric materials operating at high temperatures with very low lattice thermal conductivities.
In addition, the principal mechanism described here works for a wide class of complex disordered systems such as  structural glasses involving weakly bound atoms in void structures. 
These are in some respects similar to those of clathrates with rattling guest atom.
Finally, we note that large-scale MD simulations of heat transfer for our systems are effective to demonstrate the validity of the present theory.

\textit{Acknowledgments}~
This work is supported by the National Natural Science Foundation of China Grant No. 11334007.
T.N. acknowledges the support from Grand-in-Aid for Scientific Research from the MEXT in Japan, Grand No.26400381. J.Z. is supported by the program for Professor of Special Appointment
(Eastern Scholar) at Shanghai Institutions of Higher Learning No. TP2014012.
J. C. thanks the support from the National Natural Science Foundation of China (Grant No. 51506153).


\end{document}